\documentclass[useAMS,usenatbib,usegraphicx,onecolumn]{mn2e}
\usepackage{times}
\usepackage{graphicx}
\usepackage{amssymb}
\usepackage{multirow}
\usepackage{color}

\input{epsf}

\newif\ifAMStwofonts
\AMStwofontstrue

\makeatletter
\newcommand{\rmnum}[1]{\romannumeral #1}
\newcommand{\Rmnum}[1]{\expandafter\@slowromancap\romannumeral #1@}
\makeatother

\title[The annular gap model for young and millisecond pulsars]{The
  annular gap model for gamma-ray emission from young and millisecond
  pulsars}

\author[Y. J. Du, G. J. Qiao, J. L. Han,  K. J. Lee and R. X. Xu]
{Y. J. Du\,$^{1}$\thanks{E-mail: dyj@nao.cas.cn.},
 G. J. Qiao\,$^{2}$, 
 J. L. Han\,$^{1}$, 
 K. J. Lee\,$^{2, 3, 4}$ and
 R. X. Xu\,$^{2}$   \\
$^{1}$National Astronomical Observatories, Chinese Academy of
 Sciences, Jia-20, Datun Road, Chaoyang District, Beijing 100012,
 China \\
$^{2}$Department of Astronomy, Peking University, Beijing 100871,
 China \\
$^{3}$The University of Manchester, School of Physics and Astronomy,
 Jodrell Bank Centre for Astrophysics, Alan Turing Building,
 Manchester, M13 9PL, UK \\
$^{4}$Max-Planck-Institut f$\ddot{u}$r Radioastronomie, Auf dem
 H$\ddot{u}$gel 69, 53121 Bonn, Germany }

\pagerange{\pageref{firstpage}--\pageref{lastpage}}

\begin{document}

\maketitle
\label{firstpage}

\begin{abstract}
Pulsed high energy radiation from pulsars is not yet completely
understood. In this paper, we use the 3D self-consistent annular gap
model to study light curves for both young and millisecond pulsars
observed by the Fermi Gamma-ray Space Telescope.
The annular gap can generate high energy emission for short-period
pulsars. The annular gap regions are so large that they have enough
electric potential drop to accelerate charged particles to produce
gamma-ray photons. For young pulsars, the emission region is from the
neutron star surface to about half of the light cylinder radius, and
the peak emissivity is in the vicinity of the null charge surface. The
emission region for the millisecond pulsars is located much lower than
that of the young pulsars. The higher energy $\gamma$-ray emission
comes from higher altitudes in the magnetosphere.
We present the simulated light curves for three young pulsars (the
Crab, the Vela, the Geminga) and three millisecond pulsars (PSR
J0030+0451, PSR J0218+4232, PSR J0437-3715) using the annular gap
model. Our simulations can reproduce the main properties of observed
light curves.

\end{abstract}


\section{Introduction}

Pulsars are fascinating astronomical objects in the universe. After
more than 40 years since the discovery of the first pulsars, their
pulsed non-thermal emission has not been completely understood due to
the insufficient knowledge about the global acceleration electric
field and particle dynamics in the magnetosphere.

High energy emission (e.g. $\gamma$-ray emission) from pulsars takes
away a significant fraction of the rotational energy. Several space
telescopes were used to observe high energy emission from pulsars. Six
$\gamma$-ray pulsars and three candidates were discovered by the
Energetic Gamma Ray Experiment Telescope (EGRET) (Thompson et
al. 1999). One young pulsar (PSR B1509-58) was detected only up to 10
MeV by the Imaging Compton Telescope (COMPTEL) (Thompson
2001). Pulsars detected by EGRET show double peaks with bridge
emission.  Thanks to the launching of Astro-rivelatore Gamma a
Immagini LEggero (AGILE) and Fermi Gamma-ray Space Telescope (FGST),
more than forty new $\gamma$-ray pulsars have been discovered in the
last year, including gamma-ray only pulsars and a new population of
millisecond pulsars (Pellizzoni et al. 2009, Abdo et al. 2009b, Abdo
et al. 2009c).
%
With these data, we have good opportunities to study the open
questions, e.g., the location of the emission zones, possible particle
dynamics in the magnetosphere and so on. Theory on non-thermal high
energy emission from pulsars should be significantly improved in the
coming years.

Physical and geometric magnetosphere models have been proposed to
explain pulsar's $\gamma$-ray radiation, as we will summarize below,
differing on the acceleration region of the primary particles and the
mechanism for the production of the high energy photons.

The first is the polar cap model (Daugherty \& Harding 1982, Daugherty
\& Harding 1994, Daugherty \& Harding 1996). The acceleration region
is located in the vicinity of neutron star surface up to tens of
kilometers near the magnetic pole. The observed phase-resolved spectra
of $\gamma$-ray pulsars can also be modeled. This model favors a small
inclination angle, e.g., a nearly aligned rotator. The $\gamma$-ray
emission region is so close to the neutron star surface that high
energy $\gamma$-ray photons could be absorbed in the strong magnetic
field (Lee et al. 2010).

The polar cap model was combined with the slot gap model recently to
explain the pulsar $\gamma$-ray radiation (Muslimov \& Harding 2003,
Muslimov \& Harding 2004). A 3D model of optical-to-$\gamma$-ray
emission from the slot gap was developed to study the Crab pulsar's
light curve, phase-averaged and phase-resolved spectrum (Harding et
al. 2008). Hirotani (2008) demonstrated that the slot gap model
reproduces at most 20\% of the observed fluxes because of the small
trans-field thickness. The two-pole caustic model is proposed by
Dyks \& Rudak (2003). The gap is thin, confined to the surface of the
last open field lines, and extends from both polar caps to the light
cylinder. Double peak light curve with a large peak separation (e.g.,
Vela) can be well reproduced by this model. But this model has
difficulties in explaining the light curves with small peak
separations (e.g., B1706-44 and B1055-52).
Frackowiak \& Rudak (2005) presented spectral and light curve
properties of gamma radiation obtained by numerical modeling of some
pulsars, i.e. PSR J0218+4232, PSR J0437-4715 and PSR B1821-24.

The outer gap model is another excellent model to interpret
$\gamma$-ray emission from pulsars (Cheng, Ho \& Ruderman 1986a,
Cheng, Ho \& Ruderman 1986b, Chiang \& Romani 1992, Romani \&
Yadigaroglu 1995, Romani 1996, Zhang \& Cheng 1997, Cheng, Ruderman \&
Zhang 2000, Zhang et al. 2004, Lin \& Zhang 2009). The classical outer
gap starts at the null charge surface (inner boundary), ends at the
light cylinder (outer boundary); it is further bounded by the last
open field line (lower boundary) and a layer of electric current
(upper boundary). However, Hirotani et al. (2003) argued that the
position of the inner boundary could be shifted towards the neutron
star surface because a current at nearly the Goldreich-Julian rate
(Goldreich \& Julian 1969) is injected to the outer gap. Tang et
al. (2008) used this modified outer gap to study the multifrequency
phase-resolved spectra of the Crab pulsar.

Watters et al. (2009) have simulated a population of young pulsars and
computed the beaming pattern and light curves for the three models:
the polar cap model, the two-pole caustic (slot gap) model and the outer
gap model.
Venter et al. (2009) presented light curves of millisecond pulsars
from 3D emission modeling, in the geometric context of the polar cap,
the outer gap, and the two-pole caustic models. They found that most
of the light curves are best fit by the two-pole caustic and the outer
gap models, which indicates the presence of narrow accelerating gaps
limited by robust pair production -- even in these pulsars with very
low spin-down luminosities.

Force-free relativistic MHD were used to study the time-dependent
evolution and dynamics of pulsar magnetospheres for either aligned or
oblique magnetic geometries (Contopoulos et al. 1999, Komissarov 2006,
Spitkovsky 2006, McKinnqey 2006, Timokhin 2006 and Gruzinov 2007).
Recent efforts on theoretical understanding the high energy emission
from pulsars can be found from Bai \& Spitkovsky (2009a; 2009b). They
modeled gamma-ray pulsar light curves for the two-pole caustic model
(slot gap model), the outer gap model and the ``separatrix layer"
model using the more realistic magnetic field taken from 3D force-free
magnetospheric simulations. Their separatrix layer model might be
associated with the current sheet beyong the light cylinder. Their
simulated results indicate that the separatrix layer model can best
reproduce the observations.

The annular gap model we are developing is originally proposed by Qiao
et al. (2004) and Qiao et al. (2007). The gap region is located
between the critical field lines\footnote{ The critical field line is
  across the intersection of the null charge surface and light
  cylinder.} and last open field lines, and extends from the neutron
star surface to the light cylinder. The region for high energy
emission in the annular gap model is concentrated in the vicinity of
the null charge surface, i.e., an intermediate emission height,
different from the outer gap model.  The annular gap has a sufficient
thickness of trans-field lines and a wide altitude range for particle
acceleration.  The role of the annular gap depends on the mono-polar
voltage in the annular region, so it is favorable for short period
pulsars.  This model combines the advantages of the polar gap, the
slot gap and the outer gap models, and works well for pulsars with
short spin periods. It is a promising model to explain high energy
emission from young and millisecond pulsars.
%

In this paper, we focus on the $\gamma$-ray light curves for young
and millisecond pulsars in the framework of the annular gap model.
In \S\,2, detailed physics and geometry of the annular gap are
introduced. The results of simulated photon sky-maps and light curves
for young and millisecond pulsars are presented in \S\,3. Discussions
and conclusions are given in \S\,4.

\section{The Annular Gap Model: Physics and Geometry}

The open field line region of a pulsar magnetosphere can be divided
into two parts by the critical field lines.  One is the annular
region, which is between critical field lines and last open field
lines. The other is the core region, which is within the critical
field lines. Taking an anti-parallel rotator as an example, the radii
of the core polar region ($r_{\rm core}$) and the full polar cap
region ($r_{\rm p}$) are $r_{\rm core} = (2/3)^{3/4}R(\Omega
R/c)^{1/2}$ and $r_{\rm p} = R(\Omega R/c)^{1/2}$, respectively
(Ruderman \& Sutherland 1975), here $R$ is the pulsar's radius,
$\Omega$ is the angular spin frequency. The radius of the annular
polar region is $r_{\rm ann}=r_{\rm p}-r_{\rm core} = 0.26 R(\Omega
R/c)^{1/2}$. If pulsar's spin period is smaller, the annular radius
($r_{\rm ann}$) is larger. Therefore, the annular acceleration region
is more important for those pulsars with a short spin period (e.g.,
millisecond and young pulsars) and is negligible for older pulsars
with a large spin period.

If pulsars are bare quark stars (Xu 2002, 2005), then two kinds of
acceleration regions could be formed, namely, the inner vacuum core
gap and inner vacuum annular gap.
If pulsars are neutron stars with enough surface binding energy, the
inner vacuum core gap (Ruderman \& Sutherland 1975) can be formed for
an inclined rotator with a magnetic inclination angle $\alpha >90^{\rm
  \circ}$; while the inner vacuum annular gap (Qiao et al. 2004) can
be formed only for an inclined rotator with a magnetic inclination
angle $\alpha <90^{\rm \circ}$.
If the binding energy of a neutron star's surface is not high
enough, both negative and positive charges will flow out freely
from the surface. Then two kinds of acceleration
regions (annular and core) could be formed.
Particle acceleration is more effective in the annular gap region. The
annular acceleration region extends from the pulsar surface to the
null charge surface or even beyond it. This leads to a fan-beam
$\gamma$-ray emission, which is suitable to interpret the observed
light curves and the broad-band emission (Qiao et al. 2007).

\subsection{Coordinates for the Annular Gap}

\begin{figure}
\vspace{6.5cm}
\includegraphics{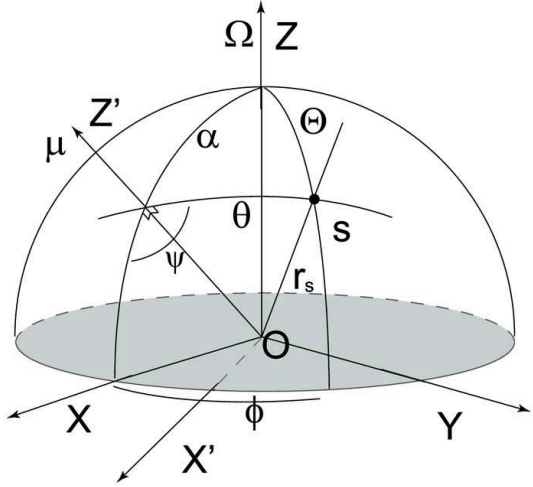}
\caption{The coordinate systems for an oblique rotator. In the
  laboratory frame $O-XYZ$, the $Z$ axis is aligned with the
  rotational axis $\rm \Omega$} of a pulsar. The magnetic frame
  $O-X'YZ'$ is generated by rotating the $O-XYZ$ coordinates around
  $Y$ axis by an inclination angle of $\alpha$, the $Z'$ axis is
  aligned with the dipolar magnetic moment {\bf$\rm \mu$}. Point ``s''
  is an emitting source on an arbitrary magnetic field line. The
  $r_{\rm s}$ is the altitude of the radiation source from the neutron
  star center. The polar angle and azimuthal angle in laboratory
  polar coordinates are denoted as $\Theta, \phi$, while the polar
  angle and azimuthal angle in the magnetic polar coordinates are
  $\theta,\psi$.  
\label{fig1}
\end{figure}
As shown in Fig.1, in the coordinate $O-XYZ$ of the laboratory frame,
the $Z$ axis is aligned with the rotational axis $\rm \Omega$ of
the pulsar. The magnetic frame $O-X'YZ'$ is generated by rotating the
$O-XYZ$ coordinate around $Y$ axis by the inclination angle $\alpha$,
the $Z'$ axis is aligned with the dipolar magnetic moment ${\bf\rm
  \mu}$. The two vectors {\bf $\rm \Omega$} and {\bf $\rm \mu$}
locates in the plane $O-XZ$, which is called the $\Omega-\mu$
plane. The polar coordinate associated with $O-XYZ$ and $O-X'YZ'$ are
called the laboratory polar coordinate and the magnetic polar
coordinate, respectively. The polar angle and azimuthal angle in the
laboratory polar coordinate are denoted as $\Theta, \phi$, while the
polar angle and azimuthal angle in the magnetic polar coordinate as
$\theta,\psi$. Similar to Lee et al. (2010), we use bold type to label
the vector or the matrix; while we use subscripts $_{x,y,z}$ and
$_{x', y', z'}$ to indicate their components in the laboratory frame
and the magnetic coordinate, respectively.

\subsection{Geometry, emission region and modeling}

We assume that pulsars have a dipole magnetic field in the magnetic
nonrotating frame. Thus dipole field line function can be expressed as
\begin{equation}
r=R_{\rm e}\sin^2{\theta},
\end{equation}
where $r$ is the polar radius from pulsar center to an emission point;
$R_{\rm e}$ is the maximum radius of the field line, which is the
function of $\alpha$ and $\psi_{\rm s}$.  To obtain the field line
constant $R_{\rm e}$ for the last open field line, Zhang et al. (2007)
and Lee et al. (2009) derived a cubic equation, i.e.,
\begin{equation}
A\cot^{3}\theta_{M}+B\cot^{2}\theta_{M}+C\cot\theta_{M}+D=0,
\end{equation}
where $A=4\sin^{2}\alpha$, $B=5\sin2\alpha\cos\psi_{\rm s}$,
$C=4\sin^{2}\alpha-6(\sin^{2}\alpha\cos^{2}\psi_{\rm
  s}-\cos^{2}\alpha)$, $D=-\sin2\alpha\cos\psi_{\rm s}$, $\theta_{M}$
is the polar angle at point M. Therefore $\cot\theta_{M}$ can be
determined analytically, and $R_{e}$ is
\begin{equation}
R_{e}(\alpha, \psi_{\rm
  s})=\frac{R_{LC}(1+\cot^{2}\theta_{M})^{3/2}}{\sqrt{1+\cot^{2}\theta_{M}
    -(\cos\alpha\cot\theta_{M}-\sin\alpha\cos\psi_{\rm s})^2}},
\end{equation}
where $R_{\rm LC}$ is the radius of light cylinder. Given the values
of $R_{e}$ and $\psi_{\rm s}$, the last open magnetic field line can be
uniquely defined.
The polar angle $\theta_{\rm N}$ of the null charge surface (defined
as $ {\bf {\Omega} \cdot B }=0 $) is given by
\begin{equation}
\theta_{\rm
  N}=\frac{1}{2}\arccos({\frac{\sqrt{8\cot^2{\alpha}\sec^2{\psi_{\rm
          s}}+9} \pm \cot^2{\alpha}\sec^2{\psi_{\rm s}}}
  {3\cot^2{\alpha}\sec^2{\psi_{\rm s}}+3}}).
\end{equation}

For a critical field line, which is across the intersection of the
null charge surface and light cylinder, there is a relation between
$\Theta_{\rm N}$ and $\theta_{\rm N}$ (Gangadhara 2004, Wang et
al. 2006), i.e.
\begin{equation}
\Theta_{\rm N}=\arccos(\cos{\alpha} \cos{\theta_{\rm N}}-\sin{\alpha}
\sin{\theta_{\rm N}} \cos{\psi_{\rm s}}).
\end{equation}
Then, the critical field line constant, $R_{\rm e, N}$, is given by
\begin{equation}
R_{\rm e, N}(\alpha, \psi_{\rm
  s})=R_{\rm LC}\csc^2{\theta_{\rm N}} \csc{\Theta_{\rm N}}.
\end{equation}
The height of the null charge surface on the last open field line can
be derived, i.e. $r_{\rm N}(\psi_{\rm s})=R_{\rm e}(\alpha, \psi_{\rm
  s})\sin^2{\theta_N}$.  Qiao et al. (2007) gave an one-dimension
solution to the acceleration electric field, the true 3D solution for
the Possion equation with mixed boundary conditions is unknown.

Dyks \& Harding (2004) found that, at low altitudes, a distorted angle
of the order of $(r/R_{\rm LC})^2$ is attributed to the rotation
deflection on the local direction of the magnetic field, where $r$ is
the radial distance of the emission source. In the annular gap model,
the radiation region is concentrated on the vicinity of the null
charge surface of the last open field lines, the emission height is
well below the light cylinder radius $R_{\rm LC}$, and the sweepback
effect on open field lines could be ignored. We use the 3D vacuum
static dipolar field in this paper.
Adopting the method of the open volume coordinates (Cheng et al. 2000,
Dyks \& Harding 2004, Tang et al. 2008, Harding et al.  2008), we
calculate the polar shape of the annular gap. The inner edge and outer
edge of the annular gap region are defined as the footpoints of the
critical field lines and last open field lines. Here we adopt the
conventional wisdom and assume that the $\gamma$-ray emissivities
$I(\theta_{\rm s}, \psi_{\rm s})$ on each open field line has a
Gaussian distribution, i.e.,
\begin{equation}
I(\theta_{\rm s}, \psi_{\rm s}) = I_{\rm peak}(\theta_{\rm peak},
\psi_{\rm s}) \exp{[-\frac{(L(\theta_{\rm s}, \psi_{\rm s})-L_{\rm
        0}(\theta_{\rm peak}, \psi_{\rm s}))^2} {2\sigma^2}]},
\end{equation}
where $L(\theta_{\rm s}, \psi_{\rm s}) = \int_0 ^{\theta_{\rm s}}
\sqrt{r^2+({\rm d}r/{\rm d \theta})^2}\,\rm d \theta $ is the arc
length of the emission point on each field line counted from the
pulsar center, $\sigma$ is a bunch scale of the emission region on
each open field line in the annular gap, and $L_{\rm 0}(\theta_{\rm
  peak}, \psi_{\rm s})$ is the arc length of the peak emissivity spot
P($\theta_{\rm peak}, \psi_{\rm s}$) on this open field line. In
principle, the P($\theta_{\rm peak}, \psi_{\rm s}$) could be located
at anywhere on this field line. However, we found that our 1-D
solution to the acceleration electric potential of each open field
lines in the annular gap reaches the maximum near the null charge
surface.  The peak emissivity of the charged particles accelerated by
the magnetospheric electric field then should be located near the null
charge surface, which is also proved by our simulated light curves
(See figures in \S3).

The dominated emission region is located near the null charge surface
(Qiao et al. 2004).
The heights $r_{\rm peak}$ for the peak emissivity spot on each open
field line (with the same $\psi_{\rm s}$, different polar angle
$\theta$) in the annular gap can be written as
\begin{equation}
r_{\rm peak}(\psi_{\rm s})=\lambda \kappa r_{\rm N}(\psi_{\rm
  s})+(1-\lambda)\kappa r_{\rm N}(0),
\end{equation}
where $\kappa$ is a model parameter for the ratio of the peak emission
height with respective to the null charge surface height $r_{\rm
  N}(\psi_{\rm s})$; $\lambda$ is another model parameter, describing
the deformation of emission location from a circle (Lee et al. 2006);
$r_{\rm N}(0)$ is the height of the point with magnetic azimuthal
$\psi=0^{\rm \circ}$ on the last open field line.
Then the peak emission position P on each open field line can be
uniquely determined, i.e., $\theta_{\rm peak}=\arcsin({\sqrt{r_{\rm
      peak}/R_{\rm e, f}(\alpha, \psi_{\rm s})}})$, where $R_{\rm e,
  f}(\alpha, \psi_{\rm s})$ is the field line constant of the open
field line with $\psi_{\rm s}$.

The peak emissivity $I_{\rm peak}(\theta_{\rm peak}, \psi_{\rm s})$
for different open field lines could follow a Gaussian distribution
(Cheng et al. 2000, Dyks \& Rudak 2003), i.e.,
\begin{equation}
I_{\rm peak}(\theta_{\rm peak}, \psi_{\rm s}) = I_{\rm 0} \exp{[
    -\frac{(\theta_{\rm sp}(\psi_{\rm s})- \theta_{\rm cp}(\psi_{\rm
        s}))^2} {\sigma_{\rm peak}^2} ] },
\end{equation}
where $I_{\rm 0}$ is a scaled emissivity, $\theta_{\rm sp}$ is used to
label a field line in the pulsar annular regions, $\theta_{\rm
  cp}=(\theta_{\rm N, \psi_{\rm s}}+\theta_{\rm p, \psi_{\rm s})}/2$
is the central field line among those field lines with $\psi_{\rm s}$.
During simulations, we take the width $\sigma_{\rm peak}\sim\,0.002$
for young pulsars, and $\sim \, 0.008$ for millisecond pulsars.

To calculate the light curves for a pulsar, we divide the polar shape
of the annular gap into 31 rings and obtain the open field line with
$\psi_{\rm s}$ and $\theta_{\rm s}$. Then we calculate the
emissivities projected onto the sky $I(\theta_{\rm j}, \psi_{\rm j})$
and direction ${\bf n_{\rm B}}(\theta_{\rm j}, \psi_{\rm j})$ of the
emission spot ($\theta_{\rm j}, \psi_{\rm j}$) on each open field line
of each ring in the magnetic frame. We also take the aberration effect
into account, and use the Lorentz transformation matrix to transform
the emission direction ${\bf n_{\rm B}}(\theta_{\rm j}, \psi_{\rm j})$
to the direction ${\bf n_{\rm \nu}}(\phi_{\rm j}, \zeta_{\rm j})$ in
the lab frame (observer frame), where $\phi_{\rm j}=\arctan ({ n_{\rm
    \nu, y}}/{ n_{\rm \nu, x}})$ is the emission spot's rotation phase
with respect to the pulsar rotation axis, and $\zeta_{\rm
  j}=\arccos({n_{\rm \nu, z}}/\sqrt{{n_{\rm \nu, x}}^2 + {n_{\rm \nu,
      y}}^2 + {n_{\rm \nu, z}}^2})$ the viewing angle for a distant,
nonrotating observer (See details for aberration effect in Lee et
al. 2010).
We also add the phase shift $\delta \phi_{\rm ret} = r_{\rm
  n}\cos(\theta_{\rm \mu, j}-\theta_{\rm j})$, the first order of
equation (33) in Gangadhara (2005), caused by the retardation effect
to $\phi_{\rm j}$, i.e. $\phi_{\rm j}=\arctan ({ n_{\rm \nu, y}}/{
  n_{\rm \nu, x}})-\delta \phi_{\rm ret}$, where $r_{\rm n} = r/R_{\rm
  LC}$ is the emission radius in units of the light cylinder radius
and $\theta_{\rm \mu, j}$ is the half opening angle of the emission
beam at the emission spot ($\theta_{\rm j}, \psi_{\rm j}$).

The ``photon sky-map'', and the corresponding light curve cut by a
line of sight with a viewing angle $\zeta$ are therefore finally
derived. The application of the annular gap to young and millisecond
pulsars for high energy pulse profiles are presented in Section 3.

\section{simulated light curves for young and millisecond pulsars}



Young $\gamma$-ray pulsars are very energetic, due to their large
spin-down luminosity. The flux of gamma ray emission detected by EGRET
are very high, e.g., $10^{-10}- 10^{-8} \, \rm erg\, s^{-1}\,s^{-2}$
(Thompson et al. 1999). We use the annular gap model to simulate the
light curves for the three brightest pulsars: the Crab, the Vela and
the Geminga, respectively. The simulated light curves can reproduce
the main observation features detected by Fermi (Abdo et al. 2010b).
%
Most of the distinct light curves of millisecond pulsars detected by
Fermi (Abdo et al. 2009a) can also be explained in the annular gap
model, especially for pulse profiles of a single peak or small
separation ($\sim$ 0.2 in phase).
We have simulated three light curves of millisecond pulsars: PSR
J0030+0451, PSR J0218+4232 and PSR J0437-4715, respectively. The
relevant results will be briefly described below. The related
parameters from simulations are given in Table 1.

\begin{table*}
\raggedright 
\begin{minipage}{153mm}
\caption
{The parameters for the simulated light curves for young and
  millisecond pulsars.\label{tbl-1}}
\begin{tabular}{ccccccccc}
\hline Pulsar & $\alpha\,(^{\rm \circ})$ & $\beta\,(^{\rm \circ})$ &
$\lambda$ & $\sigma$ ($R_{\rm LC}$) & \multicolumn{4}{c}{$\kappa$}\\
\hline
Crab & 45 & 61.3 & 0.45 & 0.15 & 0.5 ($>0.1$\,GeV) & 0.56 ($>1.0$\,GeV)
 & 0.52 ($0.3-1.0$\,GeV) & 0.45 ($0.1-0.3$\,GeV) \\
Vela & 30 & 64 & 0.9 & 0.04  & 0.72 ($>0.1$\,GeV) &
0.73 ($>1.0$\,GeV) & 0.71 ($0.3-1.0$\,GeV) & 0.68 ($0.1-0.3$\,GeV) \\
Geminga & 70 & 37.8 & 0.7 & 0.22  & 0.68 ($>0.1$\,GeV) &
0.73 ($>1.0$\,GeV) & 0.7 ($0.3-1.0$\,GeV) & 0.46 ($0.1-0.3$\,GeV) \\
J0030+0451 & 35 & 52.6 & 0.5 & 0.1 & \multicolumn{4}{c}{0.52\,
  ($>0.1$\,GeV)}\\
J0218+4232 & 30 & 50 & 0.4 & 0.12 & \multicolumn{4}{c}{0.5\,
  ($>0.1$\,GeV)}\\
J0437-4715 & 35 & 55 & 0.3 & 0.05 & \multicolumn{4}{c}{0.1\,
  ($>0.1$\,GeV)}\\
\hline
\end{tabular}
\end{minipage}
\end{table*}

\subsection{The Crab pulsar}

The Crab pulsar is famous for the phase-aligned pulse profiles of
multi-wave band emission and beautiful X-ray torus.
The observed $\gamma$-ray ($>$100 MeV) features of the Crab pulsar
(Abdo et al. 2010a and Abdo et al. 2010b) are two peaks with a
separation of $\sim$ 0.4 rotation phase and a hard bridge. The
observed light curves at different bands are quite similar. We
simulated multi-band light curves for the Crab pulsar, but
presented only one band in Figure 2 (left panels). It shows most of
the observed features, especially for the phase-aligned pulse profiles
with nearly similar peak ratio (P2/P1). An important result is that
the $\gamma$-ray emission with higher photon energy comes from higher
altitudes in the magnetosphere. This is probably caused by strong
$\gamma$-B absorption effect for emission from lower altitudes (Lee et
al. 2010).
In addition, the parameters of $\alpha$ and $\zeta$ in our model are
based on the adopted value (Wang 2003, Harding et al. 2008) and the
results of X-ray torus simulation (Ng \& Romani 2008). Other best
$\kappa$ and $\sigma$ values indicate that the $\gamma$-ray photons
are emitted from a wide range of emission altitudes from the neutron
star surface to about half of the light cylinder radius.

\begin{figure}
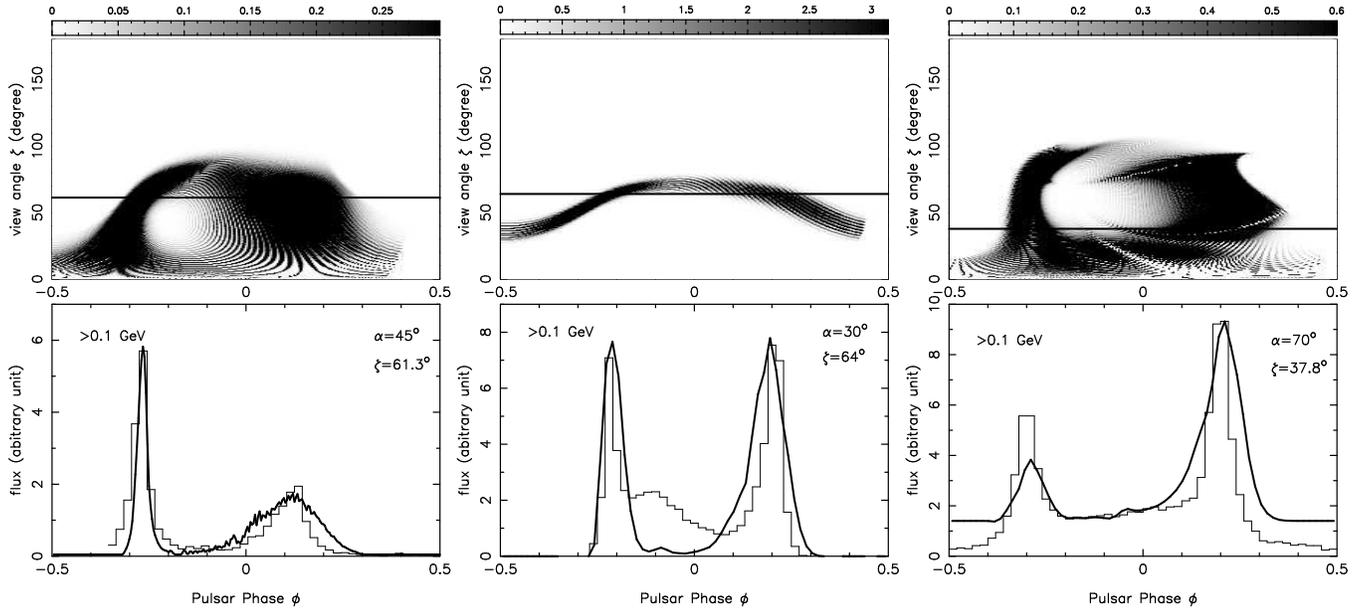

\includegraphics[angle=0,scale=.423]{f2a.eps}
\includegraphics[angle=0,scale=.423]{f2b.eps}
\includegraphics[angle=0,scale=.423]{f2c.eps}
\caption{The $\gamma$-ray photon sky-map (observer angle $\zeta$
  vs. rotation phase $\phi$) and simulated light curves for the Crab
  pulsar (left panels), the Vela pulsar (middle panels) and the
  Geminga pulsar (right panels) in the framework of single-pole
  anuular gap model, in comparison with the observations (thin solid
  lines, taken from Figure A-8, A-16 and A-12 of Abdo et
  al. 2010b). The simulated light curves are smoothed to 64 bins. The
  phase of magnetic pole in the photon sky-maps is set to be -0.15 to
  show the light curves in the middle. The observed light curves (thin
  solid lines) are compared with the simulated ones (thick solid
  lines). Parameters are given in Table 1.
  \label{fig2}}
\end{figure}

\subsection{The Vela pulsar}

Similar to the Crab pulsar, the Vela pulsar is the brightest object in
the $\gamma$-ray sky. Figure 2 (middle panels) shows the simulated
light curve of $>$ 0.1\,GeV band for the Vela pulsar. The main
features, two sharp peaks with a separation of $\sim$ 0.42 and the
peak ratio observed by Fermi and EGRET (Thompson 2001, Abdo et
al. 2010b), are approximately reproduced. Similarly, we choose the
parameters $\zeta$ according to the result of X-ray torus simulation
(Ng \& Romani 2008). The inclination angle $\alpha=30^{\rm \circ}$
which is thought to be an intermediate inclined rotator for the Vela
gives the ``best'' simulated results.  The best parameters $\kappa$
and $\sigma$ indicate that the $\gamma$-ray photons mainly come from
high altitudes. Here we only used the open field lines in the annular
gap region for the simulations. To get better results, especially for
the third peak in the bridge, both the annular gap and core gap
regions (Qiao et al. 2007) probably should be used to simulate the
observed light curves for the Vela pulsar.

\subsection{The Genminga pulsar}

The Geminga pulsar is another bright $\gamma$-ray pulsar. Figure 2
(right panels) shows the simulated light curve for Geminga. The peak
separation and peak ratio (Abdo et al. 2010b) are reproduced. The best
$\kappa$ and $\sigma$ values indicate that the emission region could
be mainly above the height of the null charge surface.



\subsection{PSR J0030+0451}

PSR J0030+0451 is a recently discovered solitary milisecond pulsar.
The pulsed $\gamma$-ray emission was detected by Fermi (Abdo et
al. 2009b).  The simulated light curve ($>$ 0.1 GeV) for PSR
J0030+0451 is shown in Figure 3 (left panels), which is similar to the
observed features.  The parameters $\alpha$ is chosen for small
magnetic inclination angles (Zhang et al. 1998, Tauris \& Manchester
1998, Young et al. 2009). Furthermore, the $\kappa$ and $\sigma$
values indicate a small emission region at low altitudes.

\begin{figure}
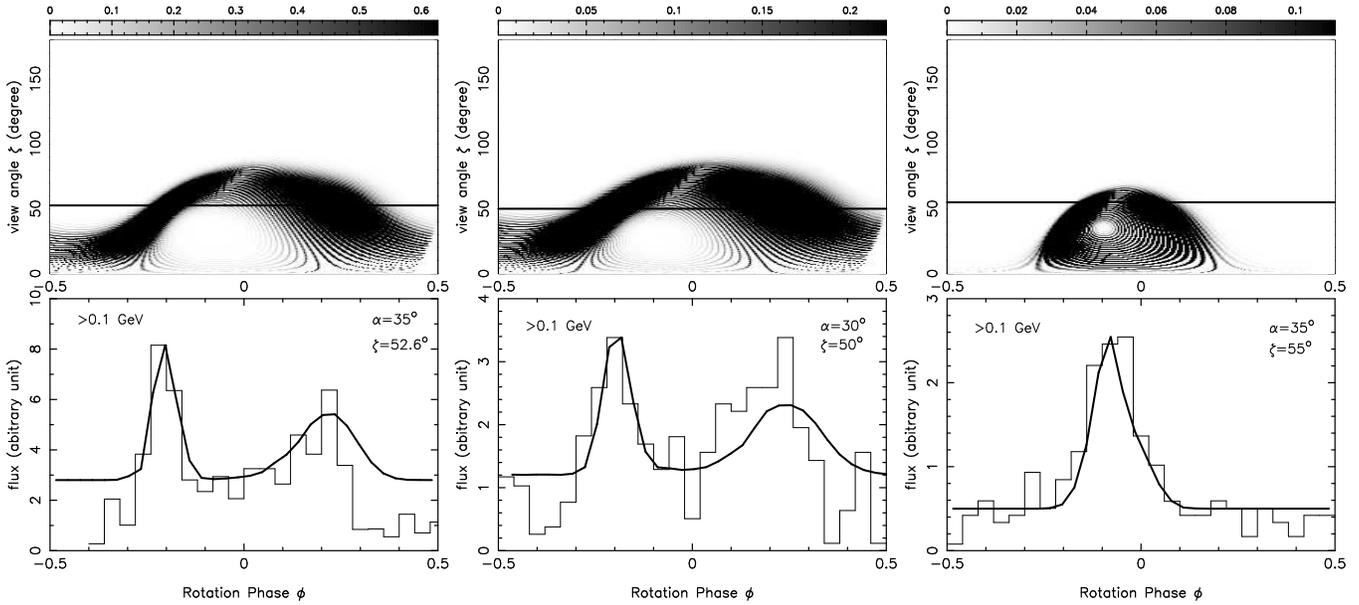

\includegraphics[angle=0,scale=.423]{f3a.eps}
\includegraphics[angle=0,scale=.423]{f3b.eps}
\includegraphics[angle=0,scale=.423]{f3c.eps}
\caption{The same as Figure 2 but for the millisecond pulsar PSR
  J0030+0451 (left panels), PSR J0218+4232 (middle panels) and PSR
  J0437-4715 (right panels) in comparison with the observations (thin
  solid lines, taken from Figure A-2, A-4 and A-7 of Abdo et
  al. 2010b, respectively). The simulated light curves (thick solid
  lines) are smoothed to 32 bins. \label{fig3}}
\end{figure}

\subsection{PSR J0218+4232}

The millisecond pulsar PSR J0218+4232 has a complex observed pulse
profile (Figure A-4 of Abdo et al. 2010b). Figure 3 (middle panels)
shows the simulated light curve ($>$ 0.1 GeV) for PSR J0218+4232. Our
result roughly reproduces the observed peaks. The model parameters,
similar to those of PSR J0030+0451, also favor a small emission region at
low altitudes.

\subsection{PSR J0437-4715}

PSR J0437-4715 is the nearest millisecond pulsar with a good radio
timing. It has a single narrow (about 0.2 phase) pulse peak at
$\gamma$-ray band (Abdo et al. 2010b). Figure 3 (right panels) shows
the simulated light curve for PSR J0437-4715. Our simulated result
reproduces the features with a small inclination angle $\alpha$ and a
large viewing angle $\zeta$. Again, the model parameters are similar
to those of PSR J0030-0451 and also favor a small emission region at
low altitudes. Note that the $\gamma$-ray emission beams (photon
sky-maps) of millisecond pulsars have a hollow cone in shape.


\section{Discussions and Conclusions}

As an approximation, a static dipole field is used to model the pulsar
magnetosphere. The pulsar magnetic field can be approximated by a
static magnetic dipole configuration if the radial distance is not so
far away from the pulsar surface (Muslimov \& Harding 2005).  The
critical magnetic field lines used to define the annular gap are
different in the magnetosphere models such as the retarded vacuum
(Cheng et al. 2000) and force-free models (Spitkovsky 2006), since the
positions of the null surface are different. However, in our model,
the $\gamma$-ray emission regions are concentrated in the middle field
lines of the annular gap, and the peak emission comes from the
vicinity of the null charge surface. The high energy emission from the
field lines near the upper boundary (critical field lines) and the
ones near the lower boundary (last open field lines) give little
contribution to the observed light curves.
Our model is different from caustic models, e.g., the outer gap model,
the slot gap model and the separatrix layer model (Bai \& Spitkovsky
2009b), which all assumed a uniform emissivity along a field line. Our
simulated light curves are mainly dependent on the non-uniform
emissivities in a deformed radiation beam (see $\lambda$ in equation
9) of the annular gap region. $\lambda$ is described in detail in Lee
et al. (2006), the large value leads to a more deformed radiation beam
from the circle one. Owing to the intermediate height of $\gamma$-ray
emission region, our annular gap model is weakly dependent on the
magnetic field configuration either static or retarded dipole field.

The annular gap has a sufficient thickness of trans-field lines and
high altitude acceleration regions. This leads to a fan-beam
$\gamma$-ray emission and sufficient photon luminosity, which is
suitable to interpret the observed light curves and the broad-band
emission. The annular gap model holds the advantages of the slot gap
and outer gap models, and works for pulsars with short spin
periods. We use the annular gap model to simulate the light curves for
young and millisecond pulsars with three assumptions: (\rmnum{1}) the
emissivities on a single field line between the critical field line
and the last open field line follows a Gaussian distribution;
(\rmnum{2}) the peak emission spot of a single field line is located
at the vicinity of the null charge surface; (\rmnum{3}) the peak
emissivities of a group of field lines with the same magnetic
azimuthal (i.e., in the same plane) between the critical field line
and the last open field line follows another Gaussian
distribution. The assumption (\rmnum{2}) is consistent with our 1-D
solution to the acceleration electric potential drop in the annular
gap. The other two assumptions are mainly based on the magnetic pair
absorption and the 3D global parallel electric field for the
acceleration of relativistic charged particles in the magnetosphere,
which will be investigated in future. In our calculations, it is shown
that the emission region extends from the neutron star surface to
about the half of the light cylinder radius, and our annular gap model
is an intermediate emission height model. We also find the following
conclusions from the modeling.


(1) The simulated light curves can reproduce most of observed features
for both young and millisecond pulsars. The $\gamma$-ray emission with
higher photon energy comes from higher altitudes in the magnetosphere.

(2) The $\gamma$-ray beams for both young and millisecond pulsars
  are hollow cones in shape.

(3) The $\gamma$-ray emission light curves (pulse profiles) are
  determined by the inclination angle of magnetic dipole field and the
  observer's viewing geometry. The magnetic inclination angles and
  viewing angles of millisecond pulsars can not well constrained by
  any methods at present. This leads to difficulties for precise
  reproduction of light curves.

(4) The radiation regions of young pulsars are larger. In terms of an
  individual field line, this large region covers from the pulsar
  surface to about the half of the light cylinder radius. The peak
  emission comes near the null charge surface.

(5) The radiation regions of millisecond pulsars are small, from the
  pulsar surface to about the one third of the light cylinder radius.
  The peak emission comes from a region below the null charge surface.

(6) Our model favors small inclination angles ($\alpha \lesssim
 35^{\rm \circ}$) for the millisecond pulsars, and larger inclination
 angles ($\alpha \sim 30^{\rm \circ} - 70^{\rm \circ}$) for the young
 pulsars. This is somewhat compatible with the alignment of the spin
 and magnetic axes over about 1 Myr from the analysis of the new
 pulsewidth data (Young et al. 2009).

(7) Our results also show that the solid angle of gamma-ray beams are
much less than $4\pi$, especially for MSPs. This will reduce the
$\gamma$-ray emission conversion efficiency $\eta=L_{\rm
  \gamma}/\dot{E}_{\rm rot}$, and solve the puzzle for $\eta > 1$ in
Abdo et al. (2010b).

As shown in Qiao et al. (2004; 2007), the annular gap can have
sufficient electric potential drop to produce pairs that can generate
radio emission. Radio emission can be generated even at a higher
region either for the annular gap or core gap.
If radio emission comes from a lower region of either inner annular
gap or the core gap (Qiao et al. 2004), the radio peak should appear
between the two $\gamma$-ray peaks. On the other hand, it is possible
that the radio radiation comes from a higher region or the opposite
magnetic pole, producing a leading or trailing radio peak.
%


\section*{Acknowledgments}
The authors are very grateful to Mr. Xue-Ning Bai and the referee for
valuable comments on the manuscript. We thank both the pulsar groups
of NAOC and of Peking University for useful conversations. Especially,
we appreciate Prof. Chou, Chih Kang for improving our
presentation. The authors are supported by NSFC (10821061, 10573002,
10778611, 10773016 and 10833003) and the Key Grant Project of Chinese
Ministry of Education (305001). K. J. Lee is also supported by ERC
Grant "LEAP", Grant Agreement Number 227947.

\end {document}